\documentclass{ws-mpla}

\begin{document}

\markboth{Francesco Cianfrani$^\dag$ and Giovanni Montani~$^\ddag$}
{A critical analysis of the cosmological implementation of Loop Quantum Gravity.}

%
\catchline{}{}{}{}{}
%

\title{A critical analysis of the cosmological implementation of Loop Quantum Gravity.
}

\author{Francesco Cianfrani}

\address{Dipartimento di Fisica, Universit\`a di Roma ``Sapienza'',\\ 
Piazzale Aldo Moro 5, 00185 Roma, Italy.\\
francesco.cianfrani@icra.it}

\author{Giovanni Montani}

\address{Dipartimento di Fisica, Universit\`a di Roma ``Sapienza'',\\ Piazzale Aldo Moro 5, 00185
Roma, Italy.\\
ENEA, Centro Ricerche Frascati, U.T. Fus. (Fus. Mag. Lab.), Via Enrico Fermi 45, 00044 Frascati, Roma, Italy.\\
INFN - Istituto Nazionale di Fisica Nucleare, Sezione di Roma 1, Piazzale Aldo Moro 5, 00185 Roma, Italy.\\
giovanni.montani@frascati.enea.it}

\maketitle

\begin{history}
\received{Day Month Year}
\revised{Day Month Year}
\end{history}

\begin{abstract}
This papers offers a critical discussion on the procedure by which Loop Quantum Cosmology (LQC) is constructed from the full Loop Quantum Gravity (LQG) theory. Revising recent issues in preserving SU(2) symmetry when quantizing the isotropic Universe, we trace a new perspective in approaching the cosmological problem within quantum geometry. The cosmological sector of LQG is reviewed and a critical point of view on LQC is presented. It is outlined how a polymer-like scale for quantum cosmology can be predicted from a proper fundamental graph underlying the homogeneous and isotropic continuous picture. However, such a minimum scale does not coincide with the choice made in LQC. Finally, the perspectives towards a consistent cosmological LQG model based on such a graph structure are discussed.
\end{abstract}

\ccode{PACS numbers: 04.60.Pp, 98.80.Qc}

\section{Introduction}

The attempts towards a quantum theory of the gravitational field find in cosmology a tantalizing puzzle for the theoretical investigation. In fact, the initial singularity makes the Universe dynamics a current issue of our understanding of the physical world. At the same time, the analysis of a cosmological space-time is simplified by space-time symmetries, which reduce the dynamical problem to the one associated with a few (finite) degrees of freedom. For these reasons, the cosmological sector of a Quantum Gravity theory stands among the main applications that any fundamental theory of interactions must address. However, one should be aware that in order to derive a symmetry reduced model, some degrees of freedom must be frozen out and this procedure conflicts with their quantum character. Therefore, a certain approximation scheme stays at the basis of any cosmological implementation and it determines to what extend the symmetry reduced model is representative of the full theory.

In this work, we will review the current status of one of the most promising Quantum Gravity approaches, Loop Quantum Gravity (LQG) \cite{revloop1,revloop2,revloop3}, with respect to the development of a proper cosmological implementation. In doing so, a critical overview on Loop Quantum Cosmology (LQC) (see \cite{revlqc1,revlqc2} for some recent reviews) will be presented and it will be sketched the realization of a model which can provide a deeper insight in the LQG cosmological sector. 
 
In LQG the quantization of the gravitational field is based on parameterizing phase space via SU(2) connections and in developing the quantum representation of the corresponding holonomy-flux algebra. Hence, the states of quantum geometry are defined on a graph structure, described combinatorially in terms of edges and vertices, with attached some SU(2) representations. It is the presence of both these two features that give geometrical operators with discrete spectra \cite{discr1,discr2,discr3} and allows to regularize the super-Hamiltonian operator \cite{hreg}. 

LQC aims to describe the cosmological sector of LQG. In this respect, the choice of variables \cite{bk} and the development of the quantum representation \cite{lqqr} follows from techniques inspired by LQG. This leads to a polymer-like description \cite{Ashpol}, whose main dynamical prediction is the replacement of the initial singularity with a bounce \cite{qnb,qnb1,qnb2}. However, in order to fix the polymer scale, at which quantum gravity effects are relevant, some external information must be added. This is due to the fact that in cosmology the spatial geometry is fixed, a part for a time-dependent conformal factor. Hence, when performing a quantization on invariant variables, as those of LQC, one lasts any kind of spatial structure, like the graph characterizing LQG states and the possibility to perform local SU(2) transformation. The discreteness of geometric operators spectra and the regularization of the super-Hamiltonian do not follow as in LQG. A consistent dynamical treatment can still be given, but an external parameter fixing the polymer scale must be introduced. As a consequence, quantum geometry is hidden behind the polymer scale. 

The development of a LQG scenario which reproduces the main features of LQC can explain how the polymer scale arises and test the viability of the proposed quantum cosmological scheme. The first attempts in this directions have been given in \cite{short}, where it has been outlined how the restriction to proper edges and surfaces within LQG implements the reduction proper of LQC variables. A similar conclusion has been inferred in \cite{impl} by analyzing the implications of the SU(2) gauge fixing associated with invariant connections. 

The result of these analysis is that a proper description of a quantum cosmological space can be derived from a fundamental graph structure made of cubical cells. Within this scheme, it has been outlined how the polymer scale can be related with the length of each edge. Indeed, there is no agreement between the polymer scale fixed in the improved dynamics of LQC \cite{qnb2} and the one fixed in \cite{short}. Therefore, the investigation of the proposed scenario in full LQG allows us to go over the LQC approach and it open up the perspective to analyze the cosmological sector of LQG.                       
The manuscript is organized as follows: in section \ref{1} the LQG framework is briefly reviewed, while in section \ref{2} LQC is presented. Then, section \ref{3} is devoted to a critical discussion on the cosmological implementation of LQG. In section \ref{4} the development of a LQG model which is able to account for cosmological symmetries is presented and the emergence of the polymer scale for LQC is demonstrated. Finally, section \ref{5} provides the outlook for such a reduced model to realize the cosmological sector of LQG. 

\section{Loop Quantum Gravity}\label{1}

LQG \cite{revloop1,revloop2,revloop3} represents the most promising approach towards a non perturbative quantum model for the gravitational field.

The loop approach towards Quantum Gravity entails that: i) phase space is parametrized by Ashtekar-Barbero connections $A^i_a$ and inverse densitized triads $E^a_i$, ii) the quantum representation of the holonomy-flux algebra is realized into the space of distributional connections \cite{ALMMT95}.  

Ashtekar-Barbero connections turn out to be proper configuration variables for gravity in the Holst formulation \cite{holst}. In fact, if one parametrizes phase space in terms of spin connections, the set of constraints is second-class, because some constraints arise besides the super-Hamiltonian, the super-momentum and the Gauss constraint of the Lorentz group. One can get rid of these additional conditions and reduce to a first-class set by fixing a proper phase space hypersurfaces. On such a hypersurfaces, $A^i_a$ and boost parameters $\chi_i$ can be taken as configuration variables. The analysis of the relic constraints outlines the emergence of the SU(2) Gauss constraint $G_i$ and the role of boost parameters, whose conjugate momenta $\pi^i$ are constrained to vanish. The Holst action can thus be written as \cite{tg} 
\begin{eqnarray}
S=\int dtd^3x\bigg[E^a_i\partial_tA^i_a+\pi^i\partial_t\chi_i-\frac{1}{\sqrt{g}g^{tt}}H+\frac{g^{ti}}{g^{tt}}H_i+\eta^iG_i+\lambda^i\pi_i\bigg],\label{FINALACT}
\end{eqnarray}

$\eta^i$ and $\lambda^i$ being Lagrangian multipliers.

Henceforth, SU(2) gauge invariance is a basic symmetry, while Ashtekar-Barbero connections capture the whole dynamical information contained in the Holst formulation for gravity. This result holds in vacuum \cite{tg} and in the presence of matter fields \cite{tgm1,tgm2,tgm3}. 

In view of SU(2) gauge invariance and back-ground independence, the variables whose algebra is quantized are not pointwise, but they are holonomies along paths $h_\alpha(A)$ and fluxes across surfaces $E_i(S)$. These object are useful because their Poisson brackets give finite expressions. Furthermore, holonomies transform at boundary points only under SU(2) transformations, while spatial diffeomorphisms act on them by pulling back the path $\alpha$. 

The development of the quantum representation starts with the enlargement of the classical configuration space to the one of distributional connections $\bar{X}$ (see \cite{revloop2} and references therein), {\it i.e.} the space of general homomorphisms $X_{l(\alpha)}$ (realized as a projective limit) from piecewise analytic paths $\alpha$ into the topological SU(2) group. $\bar{X}$ is a compact Hausdorff space in the Tychonov topology, for which a regular Borel probability measure $d\mu$ can be defined via the Haar measure of the SU(2) group. The space $\bar{A}=\mathcal{L}^2(\bar{X},d\mu)$ of square integrable functions over $\bar{X}$ is the kinematical Hilbert space \cite{ALMMT95}. 

A basis in such a space is given by spin-networks functionals $\psi_S$. Spin-networks are oriented graphs whose edges $e$ are labeled by SU(2) irreducible representations $\rho_e$ and whose vertices $v$ are labeled by intertwiners $I_v$ belonging to the tensor product of the vector spaces $V_{\rho_e}$ and $V^*_{\rho_e}$ for outcoming and incoming edges, respectively. Spin-network functionals are given by the following expression
\begin{equation}
\psi_S(g)=\prod_{v}I_v\prod_e\sqrt{2j_e+1}\rho_e(g)
\end{equation}

$j_e$ being the spin quantum number of the irreps $\rho_e$.

The essentially self-adjoint momentum operators can be defined starting from the holonomy-flux algebra.

Kinematical constraints are the SU(2) Gauss one and the supermomentum one. The former is solved by taking invariant intertwiners at vertices, while the latter is not defined in the standard treatment (however a proposal to define diffeomorphisms generators can be found in \cite{var}) and background independence is implemented defining states on knots. It is worth noting that the adopted representation of the kinematical algebra is the only cyclic one invariant under the action of spatial diffeomorphisms \cite{LOST}. 

One of the main issues of LQG is the achievement of discrete spectra for geometric operators \cite{discr1,discr2,discr3}. This is due to i)the compactness of the gauge group and ii)the existence of a fundamental combinatorial structure (edges and vertices) at a fundamental level.

Another success of LQG is the possibility to regularize the super-Hamiltonian operator \cite{hreg}.
Such a regularization implies a graph-dependent triangulation of the spatial manifold in tetrahedra. In particular, vertices $v$ are the base points for some of the tetrahedra, whose edges are developed from the segments $s_i$ starting from $v$ and belonging to a triple of edges $e_i(v)$ meeting in $v$. The expression of the super-Hamiltonian can be rewritten such that the volume, the connection and its curvature appear. The volume is a well-defined operator, while $A^i_a$ and $F^i_{ab}$ can be obtained by a limiting procedure on holonomies along an edge or a loop with decreasing length. This limit can be realized as the triangulation becomes finer and finer. Finally, one finds that as tetrahedra shrink to their base points only a finite number of terms contribute to the evaluation of the super-Hamiltonian operator and the limit is well-defined in the dual space to the diffeomorphisms invariant sector of the kinematical Hilbert space.

\section{Loop Quantum Cosmology}\label{2}

The aim of Loop Quantum Cosmology is to provide a quantum description of the Universe derived from LQG. In this respect, the symmetries proper of a cosmological model should be implemented in the LQG formulation. This point has been discussed in \cite{bk}, where they adopted invariant connections, {\it i.e.} connections invariant under the action of the symmetries proper of the considered cosmological model. This restriction breaks both SU(2) gauge invariance and background independence \cite{lqqr}.
 
For instance, in the case of a Friedman-Robertson-Walker (FRW) space-time, the 3-metric is homogeneous and isotropic. Invariant Ashtekar-Barbero connections and densitized triads read as follows in this case
\begin{equation}
A_a^i=c{}^0\!e_a^i,\qquad E_i^a=p{}^0\!E_i^a,\label{gf}
\end{equation} 

where 
$c$ and $p$ depend on time only, while ${}^0\!e^i_a$ and ${}^0\!E^a_i$ denote the so-called simplicial 1-form and densitized vectors, respectively, and they determine the homogeneous and isotropic time-independent line element. It is worth noting how the restriction to a homogeneous and isotropic spatial manifold does not fix uniquely connections and the densitized triads as in (\ref{gf}), because one can always perform a generic local rotation in tangent space without modifying the line element. Therefore, it is the choice of invariant connections which provide the SU(2) gauge symmetry breaking, rather than the restriction to a FRW space-time.  

The holonomies along edges $e$ parallel to simplicial vectors ${}^0\!e^a_i$ are given by 
\begin{equation}
h^i_e=e^{i\mu c\tau_i}, \label{hlqc}
\end{equation}

$\mu$ being the edge length, and they turns out to be linear combinations of quasi-periodic functions $N_\mu=e^{i\mu c}$. Henceforth, $N_\mu$ are taken as basis elements of the configuration space and the algebra generated by $\{N_\mu,p\}$ is quantized. The resulting Hilbert turns out to be  $\textsc{H}=\mathcal{L}^2(\textbf{R}_{Bohr},d\mu_{Bohr})$, {\it i.e.} the space of square-integrable functions over the Bohr compactification of the real line, whose measure is such that 
\begin{equation}
<N_{\mu'}|N_\mu>=\delta_{\mu',\mu}.\label{dk}
\end{equation}

The momentum operator is implemented as essentially self-adjoint and the action of phase-space coordinates on a quantum level is given by 
\begin{equation}
\hat{N}_\mu\psi(c)=e^{\frac{i\mu c}{2}}\psi(c),\qquad \hat{p}\psi(c)=-i\frac{8\pi\gamma l_P^2}{3}\frac{d}{dc}\psi(c),
\end{equation}

$l_P$ and $\gamma$ being the Planck length and the Immirzi parameter, respectively.

Having fixed all kinematical symmetries we do not have at our disposal the tools of LQG, which make the spectra of geometric operators discrete and regularize the super-Hamiltonian. However, a limiting procedure, $\mu\rightarrow0$, must be addressed in order to express the super-Hamiltonian operator in terms of quasi-periodic functions $N_\mu$. Such a limit resembles the analogous one which in the full theory provides a finer and finer triangulation. However unlike LQG, in LQC the limit does not exists. Therefore, the super-Hamiltonian is regularized ``by hand'' fixing a minimum value $\bar{\mu}$ for $\mu$.

Once $\bar\mu$ has been fixed, the action of the super-Hamiltonian is described by a difference equation. The implementation of this scheme has been performed in the presence of a free scalar field \cite{qnb1,qnb2}, playing the role of a clock matter field. It has been found that the evolution of a semi-classical wave-packet is such that the state remains semi-classical and the mean value undergoes a contraction (going backward in time) till a certain critical energy, followed by a new phase of expansion. Therefore, the initial singularity is replaced by a bounce. The dynamics is described by the following effective equation for the scale factor $a$ \cite{gr} 
\begin{equation}
\frac{\dot{a}^2}{a^2}=\frac{8\pi G}{3}\rho\left(1-\frac{\rho}{\rho_{\mathrm{cr}}}\right),
\end{equation}

where the critical energy density $\rho_{cr}$ is given in terms of $\bar\mu$ as follows \cite{he} 
\begin{equation}
\rho_{\mathrm{cr}}=\frac{3\hbar c}{8\pi \gamma^2 l_P^2\bar\mu^2 |p|}\label{cr}.
\end{equation}

In view of the $p$ factor inside the expression (\ref{cr}), a regularization scheme in which $\bar{\mu}^2|p|$ is constant would be preferable. Otherwise the scale at which quantum effects are relevant would depend on the scalar field momentum and in principle it can be much smaller than the Planck length. Hence, the $\bar\mu$ value is usually chosen in order the operator $p^2$ to reproduce the minimum eigenvalue of the area operator in LQG when acting on $N_{\bar\mu}$, which gives 
\begin{equation}
\bar{\mu}^2|p|=2\sqrt{3}\pi\gamma l_P^2.\label{barmu}
\end{equation}

In \cite{bohe} effective equations have been derived for the case of a self-interacting or massive scalar field and significant differences with respect to the standard case have been found due to quantum back-reaction. 

The extensions to anisotropic Bianchi type I, II and IX models have been performed in \cite{ashedI,mm}, \cite{ashedII} and \cite{IX}, respectively, while the descriptions of the generic cosmological solution is still an open issue \cite{bkl}.

A current line of investigation is now devoted to infer testable predictions from LQC by studying the quantum geometrical effects on the scalar and tensor perturbations of the CMB spectrum \cite{cmb}. It is worth noting the comparison with the modifications predicted in the Wheeler-DeWitt formulation \cite{wcmb}. 

\section{Cosmology and LQG}\label{3}
The cosmological sector of LQG is still elusive. The reasons for this drawback are not only due to the current difficulties in describing the quantum dynamics of the gravitational field and the absence of a proper semi-classical limit for quantum geometry. A minisuperspace model should provide some simplifications to the whole dynamical treatment, such that some issues of the general case could be solved. 

For instance, this happens in quantum geometrodynamics. The Wheeler-DeWitt equation \cite{WdW} has no well-defined quantum counterpart, because no consistent Hilbert space has ever been found for the 3-metric in Superspace \cite{K}. Some issues are solved as soon the restriction to FRW line elements takes place. There is only one configuration variable and the associated dynamical system can be reduced that of a point-like particle. Within this scheme, it has been outlined the failure of the Wheeler-DeWitt cosmological sector in avoiding the classical singularity \cite{BI}.

The problem in pursuing a minisuperspace approximation for LQG lies in the fact that the classical restriction of degrees of freedom conflicts with the development of the kinematical Hilbert space. More precisely, it has been outlined how the main results of LQG (discrete spectra of geometrical operators and the regularization of the super-Hamiltonian) are due to the fundamental graph structure together with the compactness of the gauge group and background independence. As soon as the restriction to homogeneous variables takes place, one looses any kind of spatial structure, while local SU(2) gauge symmetry is not manifest anymore. 

Granted the tension between LQG and the minisuperspace approximation, the point of view advocated in LQC is to save the latter and to perform a quantization procedure as close as possible to LQG one. The resulting picture is that of a polymer quantization \cite{Ashpol}, in which the parameter $\bar\mu$ gives the spacing of the lattice structure in configuration space. Such a lattice scale is heuristically induced from the discrete structure proper of LQG quantum space.

In order to exploit the origin of the parameter $\bar\mu$ and to sustain the above mentioned correspondence a deeper insight into the cosmological sector of LQG should be given. 

\section{Minimum scale from quantum geometry}\label{4}

The foundation of LQC has been investigated in \cite{short,impl}, starting from the analysis of the action of fluxes in a cosmological setting. 

In \cite{impl} the restriction to edges parallel to simplicial vectors ${}^0\!e^a_i$ and to the associated holonomies $h_e^i$ has been derived from the SU(2) gauge fixing proper of LQC. It has been outlined how the most general connections and densitized triads associated with a FRW line element are given by the expressions (\ref{gf}) modulo a SU(2) gauge transformation. This feature has been taken as an indication that the SU(2) gauge invariance holds also in a symmetry-reduced model and the implications of the gauge-fixing associated with (\ref{gf}) should be analyzed.

In particular, the condition fixing $E^a_i$ according with the second relation in (\ref{gf}) reads
\begin{equation}
\chi_i=\epsilon_{ij}^{\phantom{12}k}{}^0\!e^j_a E^a_k=0.\label{chi}
\end{equation}

The Poisson brackets between $\chi_i=0$ and the Gauss constraint $G_i=0$ do not vanish ``on-shell'', thus the system of constraints becomes second-class. Hence, in order to analyze the behavior of fluxes, which are defined in unreduced phase space, Dirac brackets must be considered. 

The analysis of Dirac brackets outlines that connections do not commute anymore. This result is not surprising and reflects the need to restrict configuration space according with the first relation in (\ref{gf}). Moreover, the brackets between connections and densitized triads are modified too and, as consequence, the holonomy-flux algebra changes. In order to preserve the same algebra as in LQG, which would allow to quantize the associated reduced system, a restriction on admissible paths and surfaces must be addressed. In particular, admissible paths are those ones along simplicial vectors, whose associated holonomies are $h^i_e$ (\ref{hlqc}). As for fluxes $E_i(S)$, they have to be smeared across those surfaces $S_i$ which are orthogonal to simplicial vectors ${}^0\!e^a_i$, {\it i.e.} $E_i(S_i)$. 
 
Thanks to (\ref{gf}), $E_i(S_i)$ and $p$ are related by
\begin{equation}
\sum_i E_i(S_i)=p\sum_i\Delta_{S_i},
\end{equation}

$\Delta_{S_i}$ being the area of $S_i$ in the fiducial metric.

For the left-hand side of the expression above one finds
\begin{equation}
\sum_i[E_i(S_i),h^j_e]=8\pi\gamma G  h^j_e\tau_j,\label{E}
\end{equation}

while for the right-hand side the following relation holds
\begin{equation}
\sum_i[p\Delta_{S_i}, h^j_e]=\frac{8\pi\gamma G}{3V_0}\sum_i\Delta_{S_i}\mu h^j_e\tau_j.\label{pdelta}
\end{equation}

The expressions (\ref{E}) and (\ref{pdelta}) coincide if the following ``duality condition'' holds 
\begin{equation}
|\sum_i\Delta(S_i)\mu|=3V_0. \label{dual}
\end{equation}

Henceforth, a cosmological space-time can be described in LQG by considering a graph whose edges are parallel to simplicial vectors and taking an analogous restriction for the fluxes from which geometric quantities are evaluated. Moreover, the edge length and the area of the surfaces must satisfy the relation (\ref{dual}).  
   
Within this scheme, it is possible to define geometrical operators and to demonstrate that their action do not depend on the edge length, reflecting the local character of LQG variables, while associated spectra are discrete. This point outlines how the existence of a minimum value for $\mu$ is not related with the discrete geometric structure proper of LQG.  

As soon as the super-Hamiltonian is concerned, the structure of the graph underlying the continuous picture should be given. According with the restrictions discussed previously, the simplest choice is that of a cubical lattice with a fixed 6-valence vertex structure. On such a simple graph, the action of the super-Hamiltonian can be written as the elementary action on each vertex time the total number of vertices $N_v$, {\it i.e.} 
\begin{eqnarray}
H=-\frac{3N_v\bar\mu^3}{8\pi l_P^2\gamma^2\bar\mu^2}\hat{p}^{1/2}\hat\sin^2{\bar{\mu}c},\label{shv}
\end{eqnarray} 

This expression has the correct semi-classical limit if the following condition holds 
\begin{equation}
V_0=N_v\bar\mu^3\rightarrow \bar\mu=\left(\frac{V_0}{N_v}\right)^{1/3}.
\end{equation} 

Therefore, a fixed value for the parameter $\bar\mu$ entering the super-Hamiltonian regularization can be inferred ``{\emph a posteriori}'' from a simplified LQG scenario and it coincides with the edge length of the fundamental graph structure. Such a value is constant in the standard LQG framework because the number of vertices do not change (but see \cite{alro} for a regularization procedure which gives a super-Hamiltonian operator changing the number of vertices). Therefore, the findings of this analysis conflict with the standard LQC paradigm, where $\bar\mu$ is a function of the scale factor (\ref{barmu}). 
 
The need to refer to a full LQG model in order to fix the $\bar\mu$ parameter has been outlined in \cite{bop}, where it is suggested that cosmological quantities should be defined in local patches only. In the proposed scheme, each patch corresponds to a fiducial volume $V_0/N_v$, containing a single vertex of the fundamental graph.
 
\section{Perspectives for a self-consistent quantum cosmological model}\label{5}

In the previous section, the foundation of LQC has been investigated and the emergence of a minimum scale has been inferred from a restricted LQG model. Hence, such a model captures some kinematical features of a cosmological space-time and for this reason it is a real candidate for a symmetry-reduced LQG scenario. 

The fundamental graph is made of edges parallel to simplicial vectors only. Spatial symmetries have to be implemented via proper restrictions on SU(2) quantum numbers and admissible graphs. For instance, a homogeneous and isotropic model is expected to be represented by a cubical lattice, whose edges are labeled by the same spin quantum number. Hence, vertices are 6-valent. Within this scheme it should be discussed the role of SU(2) gauge invariance in order to determine the class of spin network functionals involved in the definition of the kinematical Hilbert space and the structure of intertwiners. 

Then, having defined all the kinematical properties, one can analyze the action of the super-Hamiltonian operator. This is the main part of the proposed investigation, because a self-consistent dynamical treatment can be given thanks to Thiemann's regularization procedure \cite{hreg}. It is expected that the restriction made on the graph structure can lead to characterize the evolution of the quantum geometry and to simplify the technical issues proper of the full theory.    

The consistency of the dynamical behavior with the minisuperspace approximation is not obvious and it will determine the ability of the proposed scheme to represent an effective description for a cosmological space-time on a quantum level. Furthermore, a new insight can be given to the initial singularity issue and it is envisaged the possibility to verify the presence of the bounce. Moreover, as far as proper semi-classical states are identified, the evolution can be discussed and the comparison with the classical dynamics will offer the possibility to test for the first time the regularization procedure of LQG. 

Finally, the extension of this scheme to anisotropic or inhomogeneous space-times would follows by relaxing the assumption of fixed quantum numbers for all edges, thus opening new perspectives for the investigation of the generic cosmological solution on a quantum level. 

The presence of a fundamental graph resembles the scenario developed in spin-foam models \cite{av,rv}, where a cosmological space-time is described via a dual space-time structure triangulated by two tetrahedra only (dipole cosmology).

\subsection*{Acknowledgements}

The work of F.C. has been supported in part by the European Science Foundation, received in the framework of the Research Networking Programme on ``Quantum Geometry and Quantum Gravity'' in the form of an exchange visit grant.


\begin{thebibliography}{99}
\footnotesize\itemsep=0pt

\bibitem{revloop1}
Rovelli C., 
Quantum gravity, Cambridge University Press, Cambridge, 2004.

\bibitem{revloop2} 
Thiemann T., 
Modern Canonical Quantum General Relativity, Cambridge University Press, Cambridge, 2006.

\bibitem{revloop3} 
F. Cianfrani, O.M. Lecian, G. Montani, 
Fundamentals and recent developments in non-perturbative canonical Quantum Gravity, arXiv:0805.2503.

\bibitem{revlqc1}
Bojowald M., 
Loop Quantum Gravity and Cosmology: A dynamical introduction, arXiv:1101.5592; 

\bibitem{revlqc2}
Ashtekar A., Singh P., 
Loop Quantum Cosmology: A Status Report, arXiv:1108.0893.

\bibitem{discr1}
Rovelli C., Smolin L., 
Discreteness of area and volume in quantum gravity, 
{\it Nucl.Phys. B}, {\bf 442}, (1995), 593--622; Erratum-ibid. {\bf 456}, (1995), 753.

\bibitem{discr2}
Ashtekar A., Lewandowski J., 
Quantum Theory of Gravity I: Area Operators, 
{\it Class. Quant. Grav.}, {\bf 14}, (1997), A55--A82.

\bibitem{discr3} 
Ashtekar A., Lewandowski J., 
Quantum Theory of Geometry II: Volume operators,
{\it Adv. Theor. Math. Phys.}, {\bf 1}, (1998), 388--429. 

\bibitem{hreg}
Thiemann T., 
Quantum Spin Dynamics (QSD),
{\it Class. Quant. Grav.}, {\bf 15}, (1998), 839--873. 

\bibitem{bk}
Bojowald M., Kastrup H.A.,
Quantum Symmetry Reduction for Diffeomorphism Invariant Theories of Connections,
{\it Class. Quant. Grav.}, {\bf 17}, (2000), 3009--3043. 

\bibitem{lqqr}
Ashtekar A., Bojowald M., Lewandowski J.,
Mathematical structure of loop quantum cosmology,
{\it Adv. Theor. Math. Phys.}, {\bf 7}, 2003, 233--268. 

\bibitem{Ashpol}
Ashtekar A., Lewandowski J., Sahlmann H.,
Polymer and Fock representations for a Scalar field,
{\it Class. Quant. Grav.}, {\bf 20}, 2003, L11--1, 
arXiv:gr-qc/0211012. 

\bibitem{qnb}
Ashtekar A., Pawlowski T., Singh P.,
Quantum Nature of the Big Bang,
{\it Phys. Rev. Lett.}, {\bf 96}, 2006, 141301, 
arXiv:gr-qc/0602086

\bibitem{qnb1}
Ashtekar A., Pawlowski T., Singh P.,
Quantum Nature of the Big Bang: An Analytical and Numerical Investigation,
{\it Phys. Rev. D}, {\bf 73}, 2006, 124038, 
arXiv:gr-qc/0604013.

\bibitem{qnb2}    
Ashtekar A., Pawlowski T., Singh P.,
Quantum Nature of the Big Bang: Improved dynamics,
{\it Phys. Rev. D}, {\bf 74}, 2006, 084003,
arXiv:gr-qc/0607039. 

\bibitem{short}
Cianfrani F., Montani G., 
Shortcomings of the big bounce derivation in loop quantum cosmology,
{\it Phys. Rev. D}, {\bf 82}, 2010, 021501,
arXiv:1010.5090. 

\bibitem{impl}
Cianfrani F., Montani G.,
Implications of the gauge-fixing in Loop Quantum Cosmology, arXiv:1104.4546. 

\bibitem{ALMMT95}
Ashtekar A., Lewandowski J., Marolf D., Mourao J., Thiemann T., 
Quantization of diffeomorphism invariant theories of connections with local degrees of freedom,
{\it J. Math. Phys.}, {\bf 36}, 1995, 6456-6493, 
arXiv:gr-qc/9504018. 

\bibitem{holst}
Holst S.,
Barbero's Hamiltonian derived from a generalized Hilbert-Palatini action,
{\it Phys. Rev. D}, {\bf 53}, 1996, 5966--5969,
arXiv:gr-qc/9511026     

\bibitem{tg}
Cianfrani F., Montani G.,
Towards Loop Quantum Gravity without the time gauge,
{\it Phys. Rev. Lett.}, {\bf 102}, 2009, 091301, 
arXiv:0811.1916. 

\bibitem{tgm1}
Cianfrani F., Montani G.,
Matter in Loop Quantum Gravity without time gauge: a non-minimally coupled scalar field,
{\it Phys. Rev. D}, {\bf 80}, 2009, 084045, 
arXiv:0904.4435.     

\bibitem{tgm2}
Cianfrani F., Montani G.,
The Immirzi parameter from an external scalar field,
{\it Phys. Rev. D}, {\bf 80}, 2009, 084040, 
arXiv:0907.1530.    

\bibitem{tgm3}
Cianfrani F., Montani G.,
Gravity in presence of fermions as a SU(2) gauge theory,
{\it Phys. Rev. D}, {\bf 81}, 2010, 044015, 
arXiv:1001.2699    

\bibitem{var}
Laddha A., Varadarajan M.,
The Diffeomorphism Constraint Operator in Loop Quantum Gravity,
arXiv:1105.0636

\bibitem{LOST}
Lewandowski J., Okolow A., Sahlmann H., Thiemann T.,
Uniqueness of diffeomorphism invariant states on holonomy-flux algebras,  
{\it Commun. Math. Phys.}, {\bf 267}, (2006), 703--733,
arXiv:gr-qc/0504147.

\bibitem{gr}
Singh P., Vandersloot K., Vereshchagin G. V., 
Non-Singular Bouncing Universes in Loop Quantum Cosmology
{\it Phys. Rev. D}, {\bf 74}, 2006, 043510,
arXiv:gr-qc/0606032.

\bibitem{he}
Haro J., Elizalde E.,
Loop cosmology: Regularization vs. quantization,
{\it Europhys. Lett.}, {\bf 89}, 2010, 69001.
 
\bibitem{bohe}
Bojowald M., Hernandez H., Skirzewski A., 
Effective equations for isotropic quantum cosmology including matter,
{\it Phys. Rev. D}, {\bf 76}, 2007, 063511,
arXiv:0706.1057.

\bibitem{ashedI}
Ashtekar A., Wilson-Ewing E.,
Loop quantum cosmology of Bianchi I models,
{\it Phys. Rev. D}, {\bf 79}, 2009, 083535, 
arXiv:0903.3397

\bibitem{mm}
Martin-Benito M., Mena Marugan G. A., Pawlowski T.,
Loop Quantization of Vacuum Bianchi I Cosmology
{\it Phys. Rev. D}, {\bf 78}, 2008, 064008, 
arXiv:0804.3157

\bibitem{ashedII}
Ashtekar A., Wilson-Ewing E.,
Loop quantum cosmology of Bianchi type II models,
{\it Phys. Rev. D}, {\bf 80}, 2009, 123532, 
arXiv:0910.1278.

\bibitem{IX}
Wilson-Ewing E.,
Loop quantum cosmology of Bianchi type IX models,
{\it Phys. Rev. D}, {\bf 82}, 2010, 043508,
arXiv:1005.5565.

\bibitem{bkl} 
Ashtekar A., Henderson A., Sloan D.,
A Hamiltonian Formulation of the BKL Conjecture,
{\it Phys. Rev. D}, {\it 83}, 2011, 084024,
arXiv:1102.3474.

\bibitem{cmb}
Bojowald M., Calcagni G., Tsujikawa S.,
Observational test of inflation in loop quantum cosmology, 
arXiv:1107.1540.

\bibitem{wcmb}
Kiefer C., Kraemer M.,
Quantum Gravitational Contributions to the CMB Anisotropy Spectrum,
arXiv:1103.4967

\bibitem{WdW}
DeWitt B., 
Quantum Theory of Gravity. I. The Canonical Theory,
{\it Phys. Rev.}, {\bf 160}, (1967), 1113--1148.  

\bibitem{K}
Kuchar K., 
Canonical method of quantization, 
C. Isham, R. Penrose, D. Sciama, eds, 'Quantum Gravity 2: A Second Oxford Symposium', Clarendon Press,
Oxford, 1981, 329--374.

\bibitem{BI}
Blyth W.F., Isham C. J.,
Quantization of a Friedmann Universe Filled with a Scalar Field,
{\it Phys. Rev. D}, {\bf 11}, 1975, 768--778. 

\bibitem{alro}
Alesci E., Rovelli C.,
A regularization of the hamiltonian constraint compatible with the spinfoam dynamics, 
arXiv:1005.0817. 

\bibitem{bop}
M. Bojowald M.,
Consistent Loop Quantum Cosmology,  
{\it Class. Quant. Grav.}, {\bf 26}, 2009, 075020, 
arXiv:0811.4129.

\bibitem{av}
Battisti M. V., Marciano A., Rovelli C., 
Triangulated Loop Quantum Cosmology: Bianchi IX and inhomogenous perturbations, 
{\it Phys. Rev., D}, {\bf 81}, 2010, 064019.
arXiv:1010.1258. 

\bibitem{rv}
Bianchi E., Rovelli C., Vidotto F.,
Towards Spinfoam Cosmology,
{\it Phys. Rev. D}, {\bf 82}, 2010, 084035, 
arXiv:1003.3483 




\end{thebibliography}
\end{document}